\documentclass[]{spie}  %>>> use for US letter paper
\usepackage{amsmath,amsfonts,amssymb}
\usepackage{graphicx}
\usepackage[table,xcdraw]{xcolor}
\usepackage[colorlinks=true, allcolors=blue]{hyperref}

\usepackage{caption}
\usepackage{subcaption}

\newcommand{\ttilde}{\raisebox{0.5ex}{\texttildelow}}

\title{First tests of a 1 megapixel near-infrared avalanche photodiode array for ultra-low background space astronomy}

\author[a]{Charles-Antoine Claveau}
\author[b]{Michael Bottom}
\author[a]{Shane Jacobson}
\author[b]{Klaus Hodapp}
\author[a]{Aidan Walk}
\author[c]{Markus Loose}
\author[d]{Ian Baker}
\author[d]{Egle Zemaityte}
\author[d]{Matthew Hicks}
\author[d]{Keith Barnes}
\author[d]{Richard Powell}
\author[e]{Ryan Bradley}
\author[e]{Eric Moore}
\affil[a]{Institute for Astronomy, University of Hawai'i at M\=anoa, Hilo, HI 96720-2700, USA}
\affil[b]{University of Hawai'i at M\=anoa, Honolulu, HI 96822, USA}
\affil[c]{Markury Scientific Inc., Thousand Oaks, CA 91361, USA}
\affil[d]{Leonardo M.W. Ltd., Southampton, S015 0LG, UK}
\affil[e]{Hawaii Aerospace Corp., Honolulu, HI 96816, USA}

\authorinfo{Further author information:\\C.A.C.: E-mail: caclav@hawaii.edu}

\begin{document} 
\maketitle

\begin{abstract}
Spectroscopy of Earth-like exoplanets and ultra-faint galaxies are priority science cases for the coming decades. Here, broadband source flux rates are measured in photons per square meter per hour, imposing extreme demands on detector performance, including dark currents lower than \mbox{1 e-/pixel/kilosecond}, read noise less than \mbox{1 e-/pixel/frame}, and large formats. There are currently no infrared detectors that meet these requirements. The University of Hawai'i and industrial partners are developing one promising technology, linear mode avalanche photodiodes (LmAPDs), using fine control over the HgCdTe bandgap structure to enable noise-free charge amplification and minimal glow.

Here we report first results of a prototype megapixel format LmAPD operated in our cryogenic testbed. At 50 Kelvin, we measure a dark current of about 3 e-/pixel/kilosecond, which is due to an intrinsic dark current consistent with zero (best estimate of 0.1 e-/pixel/kilosecond) and a ROIC glow of 0.08 e-/pixel/frame. The read noise of these devices is about 10 e-/pixel/frame at 3 volts, and decreases by 30\% with each additional volt of bias, reaching 2 e- at 8 volts. Upcoming science-grade devices are expected to substantially improve upon these figures, and address other issues uncovered during testing.
\end{abstract}

% Include a list of keywords after the abstract 

\section{INTRODUCTION}
\label{sec:intro}

The high quantum efficiency (QE), low dark current (DC), and tunable cut-off wavelength of mercury cadmium telluride (HgCdTe) makes it the leading material for astronomical infrared detectors. Superb large format arrays such as the HAWAII family, manufactured by Teledyne imaging systems, are in regular use at observatories around the world, and comprise almost all the detectors on the James Webb Space Telescope. These arrays deliver dark currents of $\sim$1 e-/pix/ksec and read noise of $\sim$10-15 e-/pix/frame, which may be reduced to \mbox{2-4 e-/pix/frame} by frame averaging. However, for photon starved science such as exoplanet imaging or faint galaxy spectroscopy, such arrays are too noisy\cite{Finger2004,Downing2008}. In particular, the read noise imposes a severe barrier, and has not been improved significantly in the three decades, as it is a fundamental limitation to the MOSFET-based source follower used in each readout pixel node.

There is currently a pressing need for extremely low noise infrared detectors, as the latest astronomical decadal survey\cite{astrodecsurvey2020} identified a 6-meter space telescope operating from UV-IR the highest priority mission, with exoplanet imaging and spectroscopy of Earth-like exoplanets as the primary science driver. For this science, typical flux rates are about 1 photon per square meter per hour in V-band, and it is well known that detector noise is the most serious obstacle for such missions \cite{Robinson2020, Lacy2019, Crill2018}, leading to a need for dark currents \textless 0.001 e-/pix/s and \mbox{read noise \textless 0.3 rms e-/pix/frame}. In the optical, EMCCDs can meet these noise requirements, but in the infrared---where most of the deep biomarker spectral signatures exist---no current sensors are suitable. Ground-based infrared instruments, particularly very high resolution spectrographs, would also benefit from such sensors.

Linear-mode avalanche photodiodes (LmAPDs) offer one potential path to overcoming this read noise barrier. In these devices, large electric fields cause signal photoelectrons to be multiplied before the read noise penalty, leading to an ``effective'' read noise that is proportionally reduced by the multiplication gain. HgCdTe LmAPDs, developed by Leonardo corporation (formerly Selex) in partnership with ESO and the University of Hawai'i, have found wide use as high speed wavefront sensors, such as in the SAPHIRA detectors.\cite{Finger2010}  The current generation of SAPHIRA arrays have demonstrated sensitivity from 0.8 to 2.5 $\mu$m with high QE (\textgreater 80\%), fast pixel response, and an avalanche gain (aka APD gain) of \textgreater 500, offering an unmatched combination of sub-electron effective read noise (as low as 0.1 rms e-) at 1kHz frame rate at convenient operating temperatures around 90-100K\cite{Finger2016,Atkinson2018_2}.

While SAPHIRA detectors can be used as science focal plane detectors, they have issues that make them unsuitable for the low-flux science cases discussed above. First, their pixel format (320 x 256) is too small for integral field unit spectroscopy, which is better matched to a 1 megapixel (eg, 1k x 1k) sensor. Second, the lack of reference pixels puts high demands on voltage and temperature stability for long exposures, where drifts in the voltage level manifest as an extra noise source. Finally, when operated at low APD gain, SAPHIRAs deliver a dark current nearly at the level required, and still likely glow limited\cite{Atkinson2017}. However, when operated at high APD gain, to minimize read noise, the large voltages cause electrons to tunnel through the band gap, which causes the effective dark current to exponentially increase. In practice, this means that SAPHIRAs cannot deliver low dark current and low read noise simultaneously.

\begin{figure}[ht]
   \begin{center}
   \begin{tabular}{c}
   \includegraphics[scale=0.25]{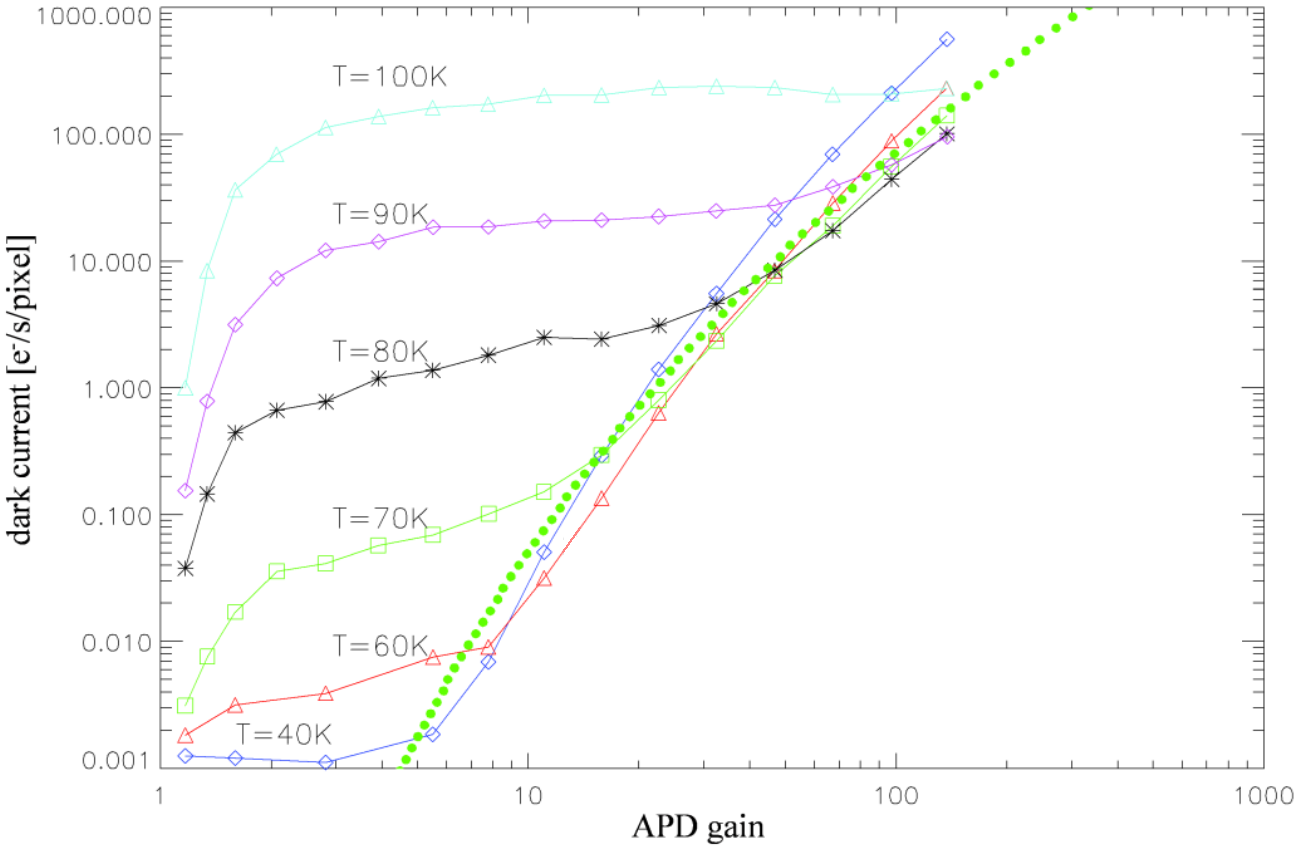}
   \end{tabular}
   \end{center}
   \caption[example] 
   { \label{fig:dc_saphira} 
Dark current versus APD gain measured for different detector temperatures (compiled from ESO[\citenum{Finger2016}] and UH[\citenum{Atkinson2017}] data). The light green curve shows the steep increase of the dark current as predicted theoretically when the tunneling current becomes dominant. The experimental data are in good agreement with these predictions. Figure adapted from Lake et al. [\citenum{lake2020developments}].}
   \end{figure}

This point does not seem to be universally appreciated, so we provide an example. Fig.~\ref{fig:dc_saphira} shows dark current measurements in a SAPHIRA device as a function of APD gain. Looking at the best 40K data (the blue curve), given a base read noise of 10 e-, for \textless 1 e- read noise, one would desire an APD gain \textgreater 10, and for \textless 0.1 e- read noise, an APD gain \textgreater 100. At these gains, the dark current increases between 1.5 and 5 orders of magnitude, respectively, which is unacceptable for low-flux science.

Motivated by the SAPHIRA results, the University of Hawai'i has partnered with Leonardo corporation, Markury Scientific, and Hawaii Aerospace to develop an LmAPD device suitable for ultra-low background infrared astronomy, with the goals of a dark current \textless 0.001 \mbox{e-/pix/s} and an effective read noise \textless 0.3 rms e-/pix/frame. The main differences from the SAPHIRA will be a larger format (1k x 1k with 15 $\mu$m pixels), a design including reference pixels to improve overall stability, and careful bandgap engineering to move the onset of tunneling current to much higher voltage, so low read noise and dark current can be simultaneously achieved. Our development plan includes maturing these sensors to technology readiness level (TRL) 5, including detailed laboratory testing in a realistic space-like environment, on-sky testing, and  characterization of performance after irradiation to space-like levels. 

This paper presents the results obtained from the engineering-grade versions of this new 1kx1k detector design. The engineering-grade detectors represent a first attempt at fabricating the megapixel detector with reference pixels. However, the engineering-grade detectors have a different bandgap structure than the science-grade design, more similar to the SAPHIRAs, so the lower tunneling current at high bias voltage cannot yet be obtained. Additionally, a contamination introduced during the manufacturing process caused the number of defective pixels to be far higher than expected, and the QE to drop at low bias voltages. Regardless, the results from the engineering grade device are encouraging, showing essentially no dark current (glow-limited) and a decreasing read noise with bias voltage in agreement with theoretical expectations.

This paper is organized as follows. Sec.~\ref{sec:detdesign} gives a detailed description of the LmAPD technology and of the design of the photodiode structure optimized to reach the performance cited above. The features of the ROIC are also discussed. Sec.~\ref{sec:labsetup} covers the testbed implemented to read out and characterize the detector. Sec.~\ref{sec:perf} presents our measurement methods of the conversion gain, read noise and dark current, as a function of the APD gain, as well as our results and their interpretation. Sec.~\ref{sec:other_effects} also reports on other effects observed with this detector. Sec.~\ref{sec:concl} gives a summary of our results and discusses the next steps planned in the development and characterization of this detector.

\section{DETECTOR DESIGN}
\label{sec:detdesign}

\subsection{HgCdTe material and avalanche process}
\label{sec:detdesign:hgcdte}

HgCdTe is unique amongst avalanche semiconductors for three reasons\cite{Baker2010,Beck2014}. First, only electrons participate in the avalanche process, because the holes have low mobility due to their high effective mass and low ionization efficiency. Second, the multiplication process is ballistic because the electrons do not experience phonon interactions or scattering, resulting in nearly noiseless amplification and deterministic APD gain as a function of bias voltage. Finally, there is no breakdown effect in HgCdTe at high bias voltages/APD gains. The signal multiplication occurs before the read noise penalty, so we speak of an ``effective'' read noise which is the read noise divided by the multiplicative gain.\footnote{For example, if a 10 e- signal is multiplied by a gain of 20 to 200 e- with an underlying read noise of 10 e-, (SNR=200/10=20), it is equivalent to a 10 e- signal seeing an effective read noise of 10e-/20 = 0.5 e- (SNR=10/0.5=20).} The penalty for this is a reduction in full well depth, which is irrelevant for low-flux applications.

The arrays are grown by Metal Organic Vapour Phase Epitaxy (MOVPE)\cite{Maxey2010}, which allows for a precise control and variable profiles of the thickness, so complex bandgaps and doping concentrations can be used to define the structure of the diode. The diode and its mesa cone form a mesa heterojunction. The MOVPE LmAPD array is hybridized to the silicon ROIC via the contact pads and indium bumps embedded at the tip of each pixel mesa cone.

\begin{figure}[ht]
   \begin{center}
   \begin{tabular}{c}
   \includegraphics[scale=0.25]{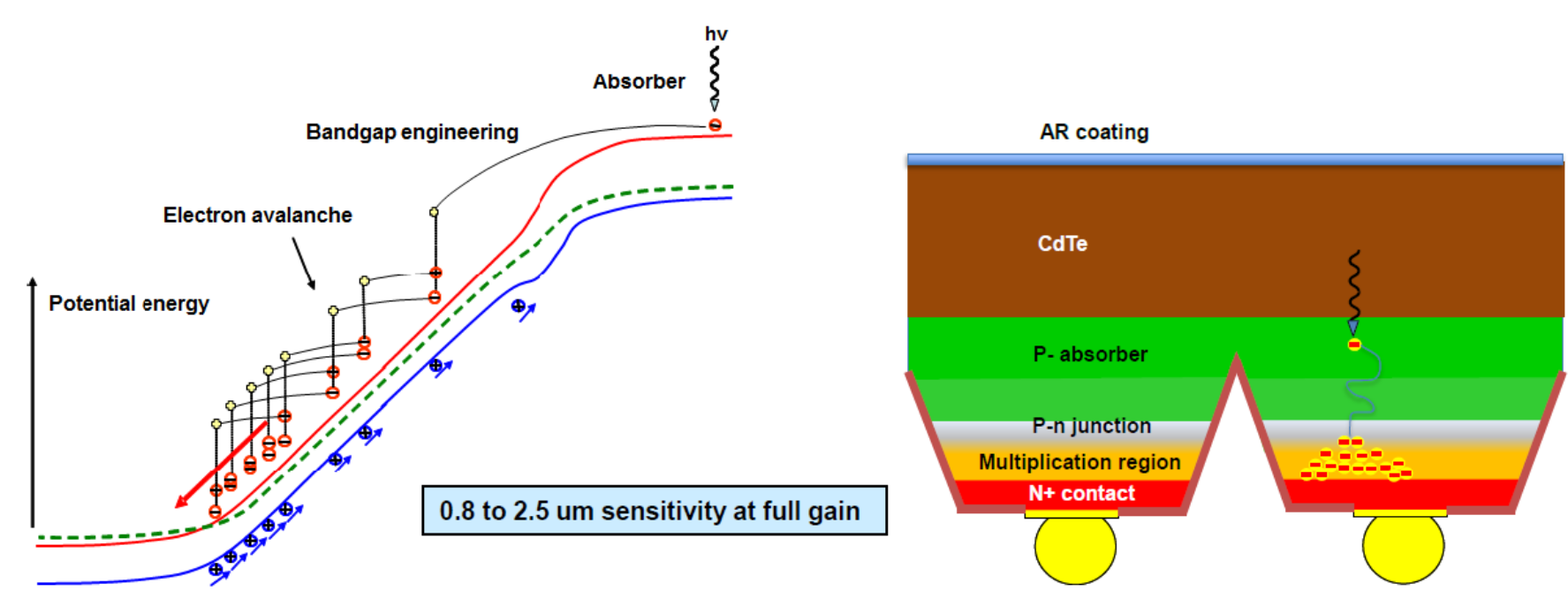}
   \end{tabular}
   \end{center}
   \caption[example] 
   { \label{fig:movpe_hgcdte_diode_design} 
The single-carrier electron avalanche process is illustrated on the left. The bias voltage is applied across the multiplication and junction regions. The right side shows a schematic of the APD heterostructure grown by MOVPE. The red circles represent the electrons experiencing the avalanche process in the semiconductor, and the blue circles represent the holes. The “yellow spheres" represent the indium bumps that are necessary to hybridize the pixel nodes to the silicon ROIC. (Original figure courtesy of Leonardo corporation)}
   \end{figure}

The top layer of the photodiode is a CdTe seed layer opaque at $\lambda $\textless0.8 $\mu$m, see Fig.~\ref{fig:movpe_hgcdte_diode_design}. Photons of \mbox{$\lambda \leqslant$2.5 $\mu$m} are absorbed in the p-type absorber directly grown on CdTe and are converted into electrons. The photon generated charge diffuses to the p-n junction and is then accelerated in the electric field of the multiplication region to start the multiplication process by impact ionization. In general, the multiplication region is made of narrow bandgap material in order to boost the APD gain relative to a given bias voltage. The absorber and the gain region are therefore decoupled so each can be optimized separately. It is possible for photons with longer wavelengths ($\lambda$\textgreater 2.5 $\mu$m) to penetrate the absorption layer and the junction to be absorbed directly in the multiplication layer.

The mesa cones on the back of the device fulfill several critical functions. Their slots extending through the absorber layer allow for complete electrical isolation between pixels to eliminate lateral charge collection. Their shape maximizes photon trapping within the pixel and prevents stray photons from scattering into neighboring pixels. This leads to near-ideal MTF (Modulation Transfer Function) and crosstalk. Together with a continuous absorber layer, this photodiode layout provides complete photon absorption within the pixel, and ensures every photoelectron receives the same APD gain independent of the position where the photon is absorbed in the pixel so that no microlenses are needed.

\subsection{Diode structure design for low dark current}
\label{sec:detdesign:dark current}

At temperatures \textgreater 60K, the temperature dependence of dark current is mainly determined by the narrow bandgap multiplication region. At high APD gains, dark current is dominated by trap-assisted tunneling current. During this process, two mechanisms are involved: an electron tunnels (tunnel-tunnel current) or is thermally excited (thermal-tunnel current) into a trap before tunneling to the conduction band. At temperatures $\leq$60K, only tunnel-tunnel current is significant. When the devices are operated at high bias voltage (and hence low read noise), this also causes the onset of tunneling current after a point, which can raise dark current to unacceptable levels. As seen in Fig.~\ref{fig:dc_saphira}, for a given temperature and bias voltage \textgreater2V, the measured dark current first exhibits a moderate increase as a function of bias voltage. Then, after the voltage threshold is crossed, the tunneling current becomes the dominant component of the dark current. It increases steeply in agreement with the exponential increase predicted by theory, as shown by the light green dotted curve.

The reference design of this detector was optimized for long integration times expected for low-flux science and has a graded bandgap structure in the multiplication region. This design should allow the thermally generated dark current to experience much lower APD gain and APD gains of several tens before the onset of trap-assisted tunneling current occurs. The engineering-grade device presented here lacks the graded bandgap.

\subsection{Characteristics of the ME1070 ROIC}
\label{sec:detdesign:roicme1070}

\begin{table}[]
\centering
\begin{tabular}{|l|ccc|}
\hline
 &
  SAPHIRA &
  \begin{tabular}[c]{@{}c@{}}ME1070\\ (engineering-grade, \\ this work)\end{tabular} &
  \begin{tabular}[c]{@{}c@{}}ME1070\\ (science-grade,\\ not yet tested)\end{tabular} \\ \hline
Pixel size                 & 24      & 15          & 15                 \\
Format                     & 320x256 & 1024x1024   & 1024x1024          \\
Max frame rate             & 1000 Hz & $\sim$10 Hz & $\sim$10 Hz        \\
Reference pixels           & No      & Yes         & Yes                \\
Video outputs              & 32      & 16          & 16                 \\
Bandgap structure          & Widened & Widened     & Widened and graded \\
Onset of tunneling current & 8 volts & 8 volts     & 15 volts (est)     \\ \hline
\end{tabular}
\caption{\label{tab:comp}Comparison of detector properties of SAPHIRA and ME1070}
\end{table}

Leonardo ME1070 1kx1k ROIC is a low-voltage multiplexer (nominal operating voltage of less than 3.5V) designed for use with avalanche photodiode arrays. A small integration node capacitance per pixel of about 27 fF is implemented to limit read noise. All of its I/O pads are on one side in order to make the arrays 3-side buttable. User-tunable options are controlled via a serial interface.

The device has 16 analog video outputs able to run at a pixel clock up to 1MHz resulting in a maximum full-frame readout speed of about 15 frames per second (FPS). The ROIC topology offers flexible readout schemes. It preserves the multiplexing advantage of 16 parallel video channels, organized in such a way that they read out 16 adjacent pixels simultaneously with a single ADC (Analog to Digital Converter) conversion strobe, so that windowing can be as small as 1 row by 16 columns. ME1070 provides scanning modes allowing for multiple digitization strobes per pixel or on a row by row basis. This facilitates multiple non destructive readouts and Fowler sampling to further minimize the read noise. The detector can be read out using the readout scheme called integrate then read (ITR) in conjunction with up-the-ramp (UTR) sampling; an external strobe to reset the full array is applied at the start of each new exposure. The readout scheme called read-reset-read for efficient correlated double sampling is also available.

The ROIC architecture has 4 reference rows which can be configured in interlaced mode. Reference pixels in NIR arrays require replacing the HgCdTe photodiode with a matching capacitor to the same bias voltage. Reference rows are commonly used to correct for settling and compensate for voltage and temperature drifts over long exposure times, which is very useful for measuring low dark current over periods of hours or days. The ROIC architecture also has 4 reference outputs coupled to a row of 256 reference pixels for further compensation, especially electromagnetic interference.

ME1070 is specifically designed for low IR on-chip self-emission circuitry. It is supplied by lower voltages and the whole top surface of the silicon ROIC has a continuous metal coverage to further exclude any glow photons. A list of properties and comparison to the earlier SAPHIRA detectors is given in Table \ref{tab:comp}.

\section{LABORATORY SETUP}
\label{sec:labsetup} 

\subsection{Cryostat and sensor chip assembly}
\label{sec:labsetup:cryostat}

\begin{figure}
\begin{subfigure}{0.45\columnwidth}
\includegraphics[width=\linewidth]{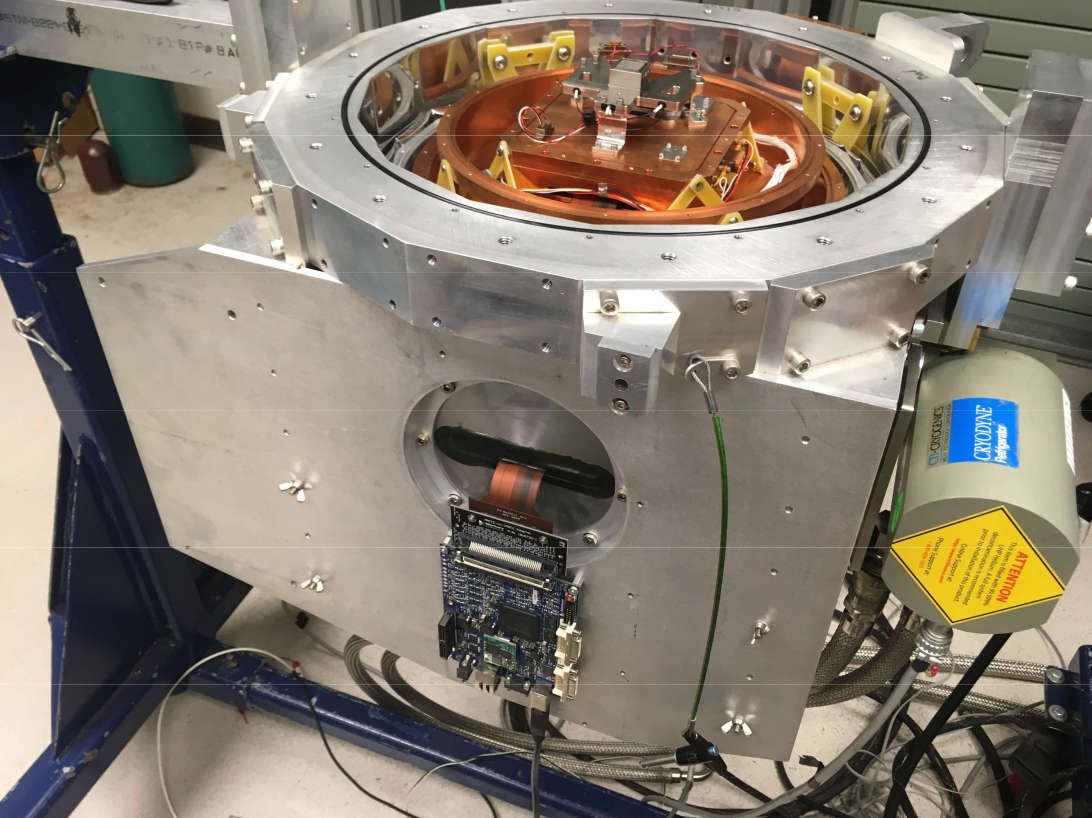}
    \caption{The test dewar with the safety cover removed, showing the detector mounting site. The MACIE card and connected flex cable are visible in the foreground. The flex cable is routed through the cryostat wall to the SIDECAR ASIC.}\label{fig:ulbcam}
\end{subfigure}\hfill
\begin{subfigure}{0.45\columnwidth}
\includegraphics[width=\linewidth]{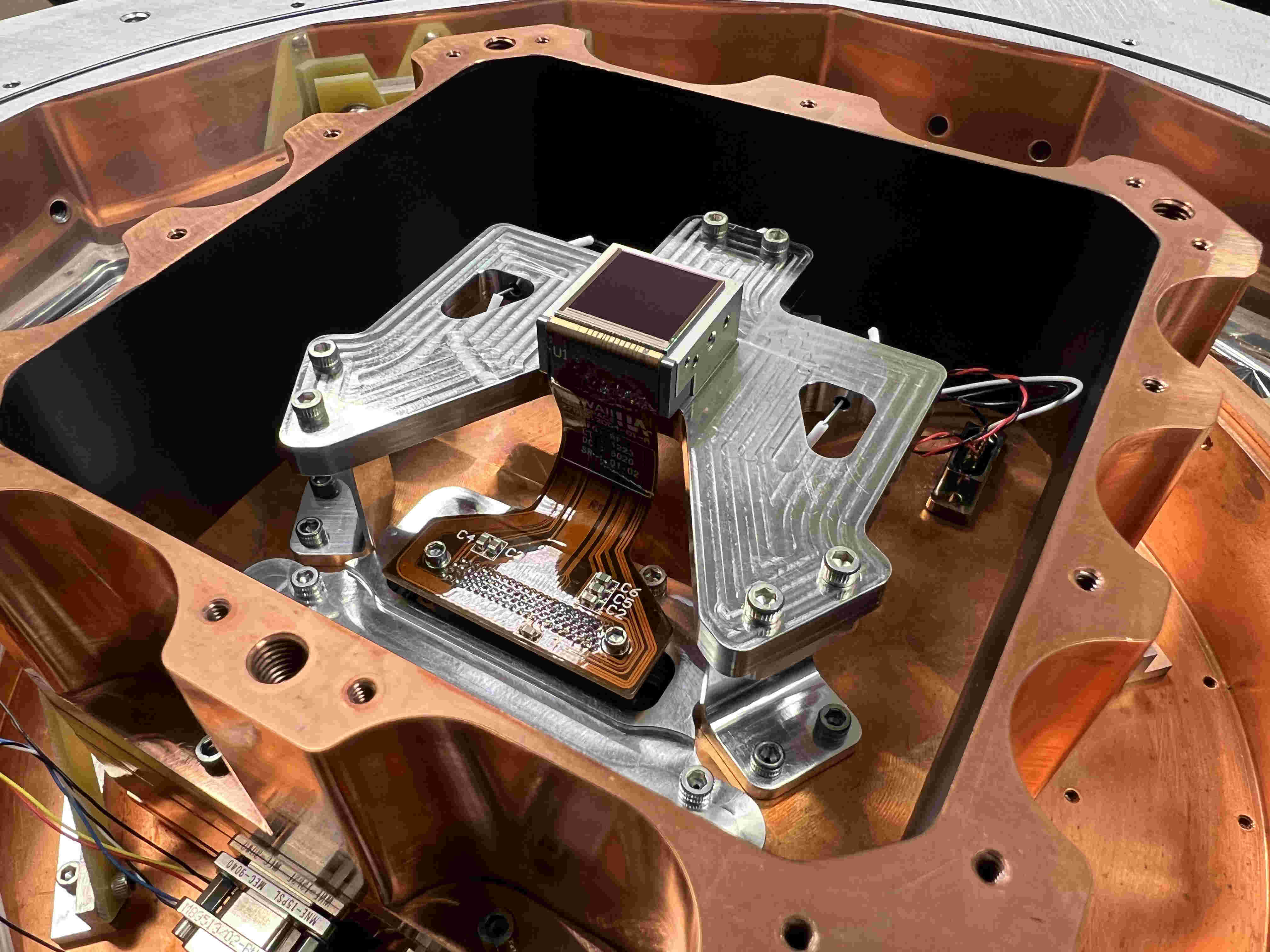}   
    \caption{The inner detector housing and mounting structures. An engineering-grade sensor is integrated on its carrier and connected to the flex cable plugged to the SIDECAR ASIC located underneath the mounting site. \newline}\label{fig:inner_det_housing}             
\end{subfigure}\\[1em]
\begin{subfigure}{0.45\columnwidth}
\includegraphics[width=\linewidth]{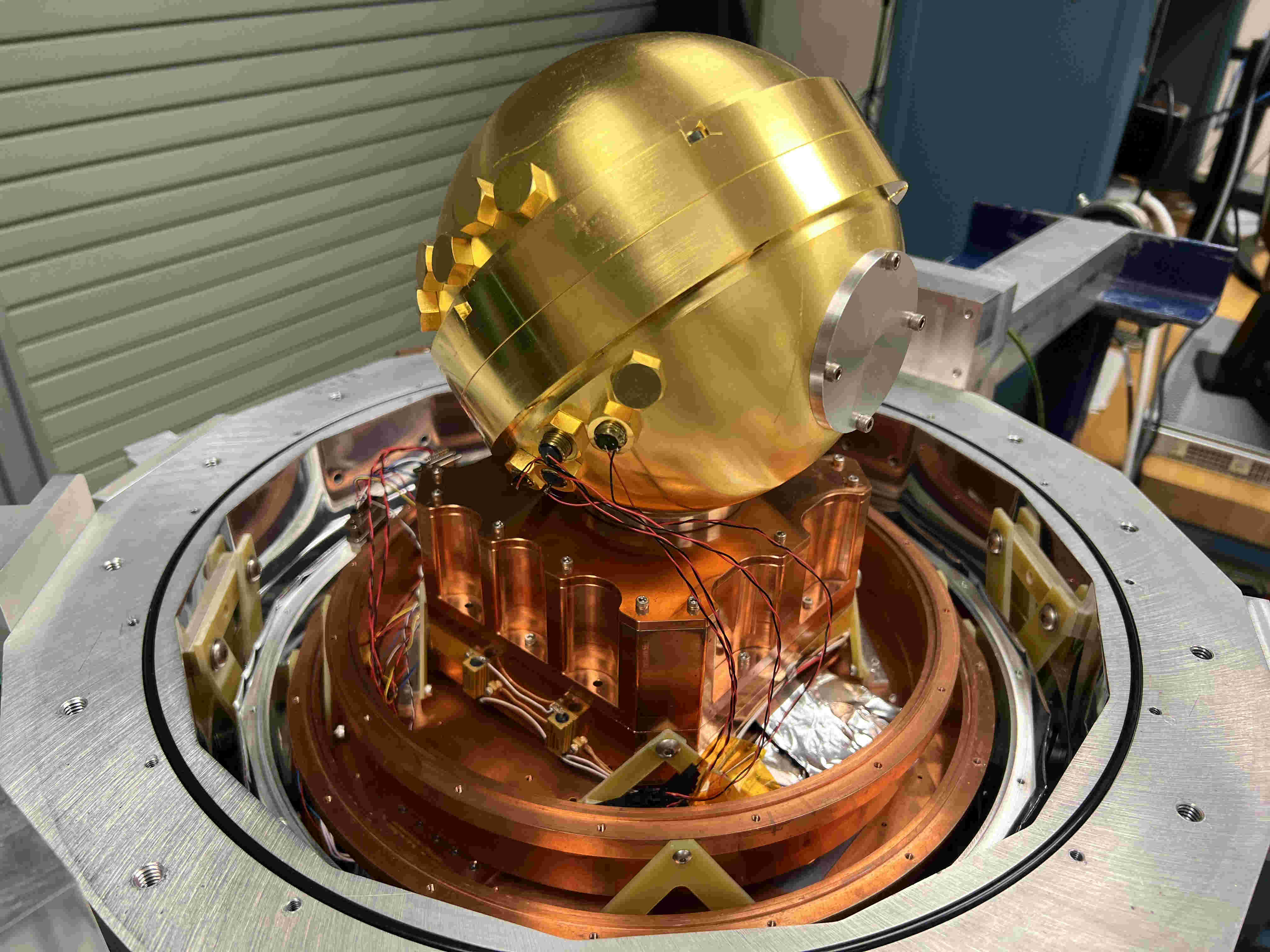}
    \caption{Gold integrating sphere mounted on top of the detector. LEDs are fitted in the ports of the integrating sphere, visible just below the central seam.\newline}\label{fig:integrating_sphere}
\end{subfigure}\hfill
\begin{subfigure}{0.45\columnwidth}
\includegraphics[width=\linewidth]{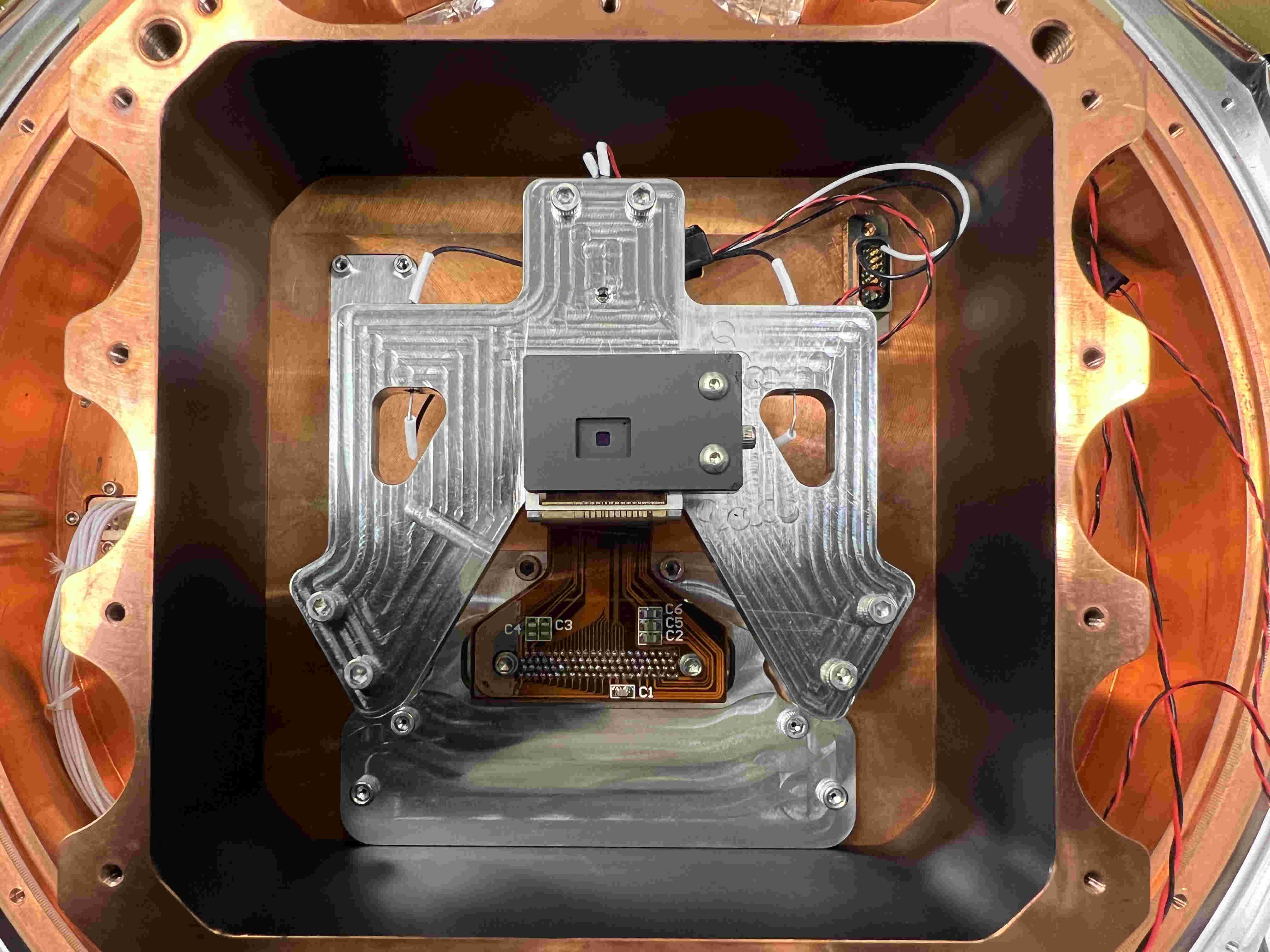} 
    \caption{The black matte mask is installed just above the detector. The square hole is visible in the  center.\newline \newline}\label{fig:det_mounting_site_w_mask}
\end{subfigure}
\caption{Pictures showing the main components of our ULBCam test dewar.}
\label{fig:testbed}
\end{figure}

The evaluation of the performance of the arrays is made using the Ultra-Low Background infrared Camera\cite{hall20044k} (ULBCam) instrument (Fig.~\ref{fig:ulbcam}), which functions as both a lab testbed and on-sky camera at the University of Hawai'i 2.2m telescope. The cryostat chamber vacuum is pumped to \textless 1e-6 Torr and can be cooled down to a temperature as low as 40K, with milliKelvin stability over week-long periods. The internal focal plane assembly is shown in Fig.~\ref{fig:inner_det_housing}, on the upper-right (see the description in the legend).

To characterize the detectors, we operate ULBCam in a testbed configuration, where a light-tight cover replaces the field lens to minimize thermal background. A cryogenic integrating sphere is placed in front of the detector, providing flat illumination with LEDs of 1.05, 1.30 (J-band) and 1.70 (H-band) $\mu$m, each with \ttilde10\% spectral bandwidth, see Fig.~\ref{fig:integrating_sphere}. It was found that the integrating sphere did not provide a perfectly light tight seal, with photoelectrons being detectable at the level of a few electrons per pixel per hour. To mitigate this, we installed a mask $\sim$1mm above the sensor, see Fig.~\ref{fig:det_mounting_site_w_mask}. This mask is at the same temperature as the detector and has minimal thermal emission in wavelengths of interest. It has a rectangular opening in the center leaving about 1\% of the sensor surface exposed. This allows measurements requiring both partial illumination and dark conditions to be made with the same system. 

\subsection{Readout chain and data acquisition}
\label{sec:labsetup:readout}

\begin{figure}[ht]
   \begin{center}
   \begin{tabular}{c}
   \includegraphics[scale=0.25]{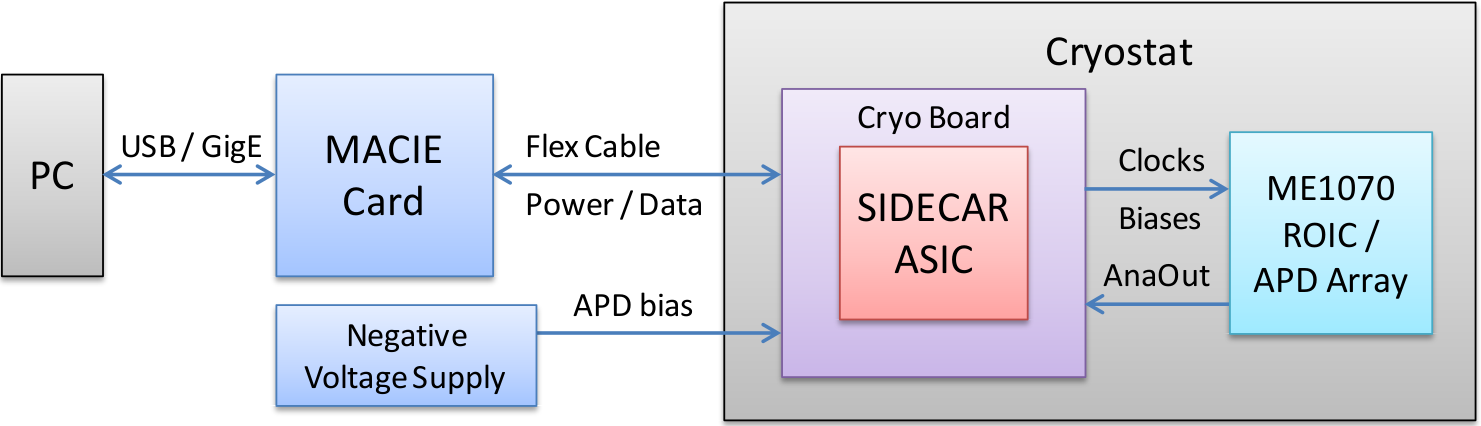}
   \end{tabular}
   \end{center}
   \caption[example] 
   { \label{fig:rdout_chain_blkdiag} 
Block diagram of the readout system for testing.}
   \end{figure}

All the hardware components of the readout chain from the detector to the computer are shown in the block diagram in Fig.~\ref{fig:rdout_chain_blkdiag}. The detector is controlled by a SIDECAR ASIC\cite{Loose2003} via an interface rigid flex cable supplied by Hawaii Aerospace. The SIDECAR is a TRL-9  microcontroller-based system-on-a-chip that comprises low noise analog circuits, detector output conditioning, and 36 parallel ADC channels for 16-bit signal digitization. It can be operated both at ambient temperature and at cryogenic temperatures down to 40K. The SIDECAR interfaces to our control computer via a MACIE (Multi-purpose ASIC Control and Interface Electronics) controller card. We developed custom firmware and software to read out the detectors via the SIDECAR and MACIE card, including a custom Python wrapper to the MACIE C library to simplify automation and testing, which we intend to release in the near future, as it is compatible with any detectors using SIDECAR or ACADIA \cite{Loose2018} ASICs. Finally, an external negative voltage supply is used to bias the detector and control the APD gain.

\begin{figure}[ht]
   \begin{center}
   \begin{tabular}{c}
   \includegraphics[scale=0.6]{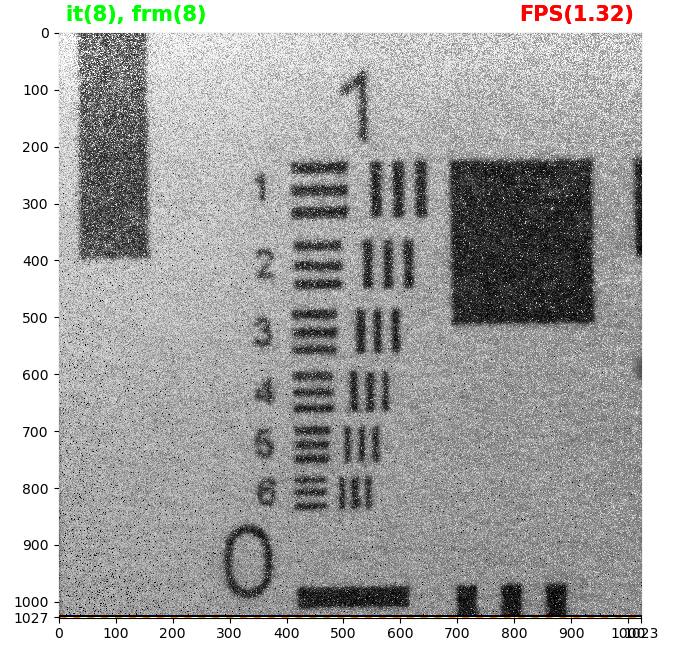}
   \end{tabular}
   \end{center}
   \caption[example] 
   { \label{fig:1st_litgh_image} 
First light image taken (2021/07/15) with our initial LmAPD engineering-grade device. A poor-quality USAF test target printed out from our office printer was imaged on the sensor. The mask over the detector was not present for this test.}
   \end{figure}

To validate our readout chain, we first operated ULBCam in the standard astronomical camera configuration with the field lens in place, then installed some foreoptics to create an image on the focal plane. Fig.~\ref{fig:1st_litgh_image} shows the first light image we took with our first engineering-grade device, showing good cosmetic quality overall.

\clearpage

\section{CONVERSION GAIN, READ NOISE \& DARK CURRENT}
\label{sec:perf}

Unless otherwise stated, all the results presented in this section were obtained with the same engineering-grade detector, cooled down to a temperature of 50K, and set up in source follower mode, with its minimum operating voltages.

\subsection{Conversion gain and read noise}
\label{sec:perf:ptc}

\begin{figure}
\begin{subfigure}{0.53\columnwidth}
\includegraphics[width=\linewidth]{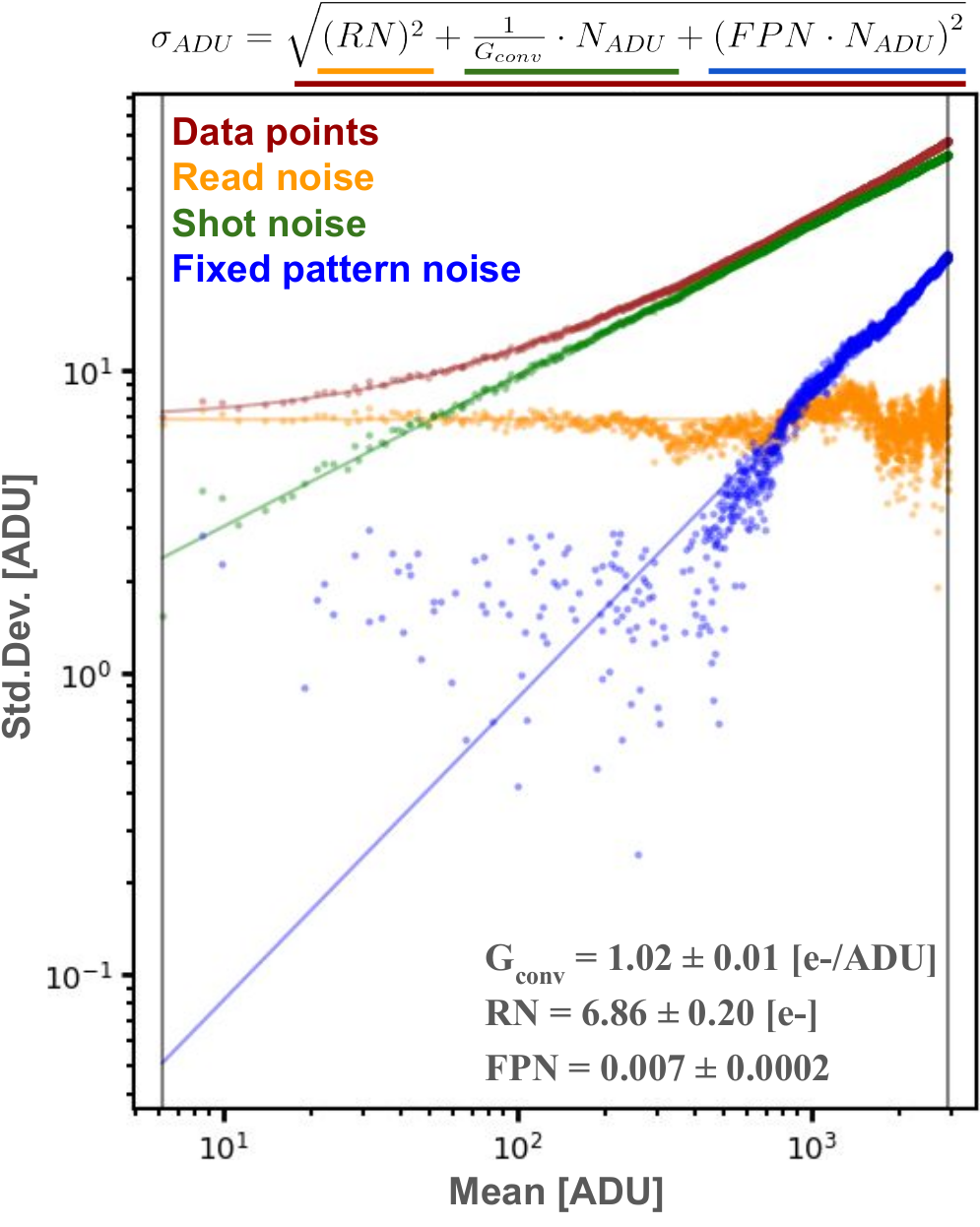}
\end{subfigure}\hfill
\begin{subfigure}{0.43\columnwidth}
\includegraphics[width=\linewidth]{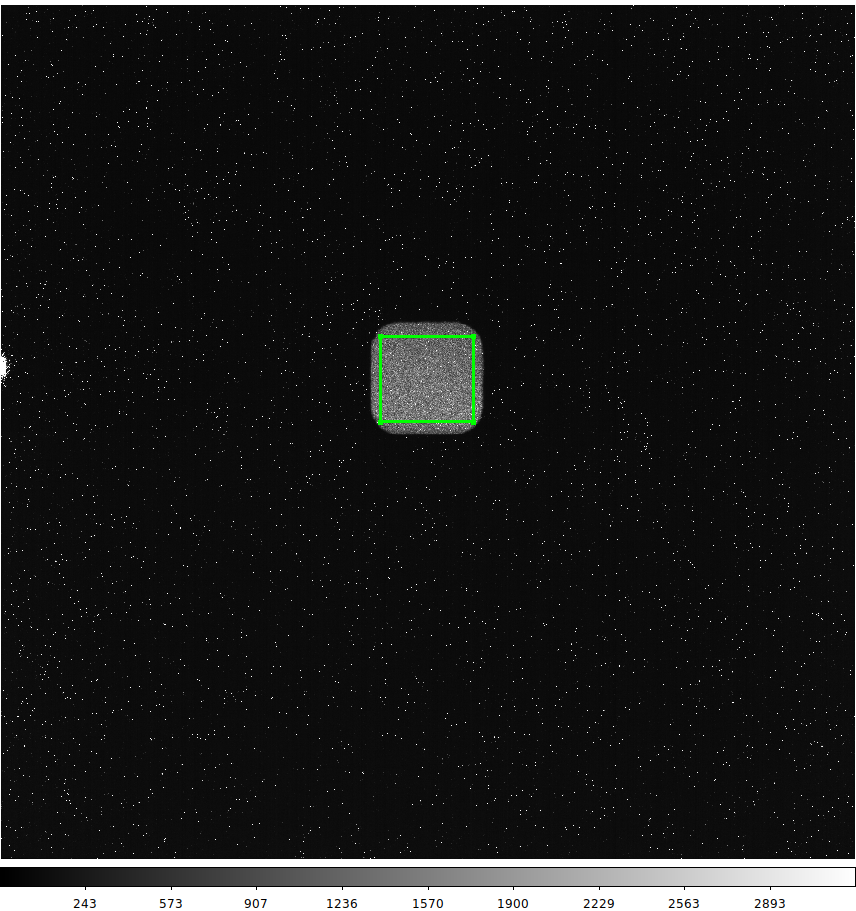}   
\end{subfigure}
\caption{On the left, example of photon transfer curve obtained for bias voltage=4.0V. The standard deviation of the signal is plotted against its mean. The signal is measured in the central portion of the sensor exposed through the window of the mask covering the sensor to the illumination output from the integrating sphere. On the right, the exact sub-region selected is indicated by the green rectangle overlaid on one of the frames included in the UTR-sampled data used for the PTC. The illumination within this sub-region of $\sim$110x110 pixels is uniform. The flat field image on the right corresponds to a data point in the PTC with an average signal of $\sim$1500 ADU. The fitted function shown in the plot title is decomposed into its three color-coded components. The values of the three target parameters ($RN$, $G_{conv}$, $FPN$) of the fitted function are also given in the legend. While the term $RN$ in the equation is in fact in [ADU], the final result given in the legend has been converted to [e-].}
\label{fig:ptc_ff}
\end{figure}

The conversion gain and the read noise are estimated using the photon transfer curve (PTC) method\cite{Janesick2007}. The standard deviation (STD) of the signal measured with the detector is plotted against the mean signal by scanning an increasing level of incident flux. The detector is illuminated by the integrating sphere with a constant flux emitted from the LED. The collected signal is sampled up the ramp, see Fig.~\ref{fig:ptc_ff}. Mean and standard deviation are computed by spatially averaging over the pixels and are expressed in analog-to-digital units (ADU). Two data cubes are produced. Each is  first subtracted from their respective bias frame in order to remove  kTC noise, then each individual frame is subtracted from the median of its four reference rows at the column level. The mean signal is computed using the sum of the two datasets frame by frame, then divided by two. The standard deviation is computed using the difference of the two datasets, then divided by $\sqrt{2}$. This last operation has the advantage of mitigating the impact on the variance of a possibly slightly non-uniform illumination. \textbf{The measured read noise is therefore defined as the CDS read noise.}

The conversion gain and read noise are obtained by fitting the function in Eq.~(\ref{eq:ptc}) to the PTC data. Eq.~(\ref{eq:ptc}) has three terms under the square root: the read noise ($RN$, here in [ADU]) corresponding to the y-intercept of the PTC, the conversion gain $G_{conv}$ corresponding to the inverse of the slope of the linear part of the PTC, and the fixed pattern noise ($FPN$) corresponding to the coefficient of the quadratic, or non-linear, part of the PTC. This last term is usually introduced to take into account that the response between individual pixels may vary. The conversion gain depends on the charge gain $g$, the APD gain $G_{avl}$, and the excess noise factor ($ENF$). As discussed in Ref.~\citenum{Finger2016}, we expect that at a temperature of 50K and relatively low bias voltages, the $ENF$ is close to 1 and constant over the considered range of bias voltages. This is verified by measuring the conversion gain as a function of bias voltage. \textbf{The normalized dark current and the effective read noise are directly derived by multiplying their raw value measured in [ADU] by the conversion gain, in order to convert them into [e-] and correct them for the APD gain.}

\begin{equation}
\label{eq:ptc}
\sigma_{[ADU]} = \sqrt{RN^2 + \tfrac{1}{G_{conv}} \cdot N_{[ADU]} + \left( FPN \cdot N_{[ADU]} \right)^{2}},\qquad \text{with}\ G_{conv} = \frac{g}{G_{avl}\cdot ENF}
\end{equation}

The PTC method can give inconsistent results when used with different sub-regions of the detector, different illuminations, etc. We developed a Gaussian process fitting method that accounts for these correlations and returns reliable uncertainties on the target parameters $\{RN, G_{conv}, FPN\}$ sampled using an MCMC sampler. At the end of the procedure, each parameter is defined by a distribution of samples. The median of the distribution is the adopted value of the parameter, and its width from the 16th to the 84th percentile is the uncertainty on the parameter. This code will be released shortly.

CMOS sensors are affected by the interpixel capacitance (IPC)\cite{Finger2005}. This comes from the coupling between neighboring pixels' capacitance when pixels accumulate charge and results in a reduction of the variance computed by simple spatial averaging. As a consequence, the linear slope of the PTC is reduced, which leads to an overestimation of the conversion gain and affects estimates of other quantities that depend on it. Even sub-percent coupling coefficient can induce an error of the order of percent on the conversion gain. We characterized the interpixel capacitance using the method described in Ref.~\citenum{Moore2006}, which recovers the correct variance using autocorrelation techniques. We derive a first measurement of the coupling coefficient of the order of 0.7\%, which leads to a systematic error of $\sim$5\% on our conversion gain. We have not corrected for this effect yet, as it is minor.

Our PTC measurements at different bias voltages are reported and discussed Sec.~\ref{sec:perf:results:perfs}.

\clearpage

\subsection{Dark current}
\label{sec:perf:results}

We measured the dark current with 2h-long integrations, sampling the detector continuously up-the-ramp at regular time intervals of 20s. In total, the resulting data cube is made up of 360 read frames. The dark current estimate is obtained by subtracting the median at the pixel level of a stack of 60 consecutive read frames taken at the end of the integration, minus the median of a stack of 60 read frames taken at the beginning of the integration, the difference being then divided by the effective integration time. The first 20 frames are discarded in order to avoid settling effects that may occur at the beginning of the acquisition. The mean dark current value reported is given by the peak of the pixel distribution and the uncertainty $\sigma$ on this value is given by the median absolute deviation of the distribution around the dark current peak. The pixels outside $\text{dark current}_{\text{peak}} \pm 3 \sigma$ are rejected as outliers. Dark current measurements at different bias voltages are reported and discussed Sec.~\ref{sec:perf:results:perfs}.

In order to examine the dependence of the measured dark current on the number of read frames included during the integration, we reproduced the method outlined in the paper Ref.~\citenum{Regan2020}. This method separates the dark current into two components: a “continuous/per-time" component which is the intrinsic dark current originating from the HgCdTe material, and a “per-frame” component which is due to the glow of the ROIC when the acquisition of a read frame is triggered. The glow contribution depends on the number of frames, while the intrinsic dark current does not. Datasets of the same total exposure time taken with two different frame times suffice to separate these quantities. 

\begin{figure}[ht]
   \begin{center}
   \begin{tabular}{c}
   \includegraphics[scale=0.6]{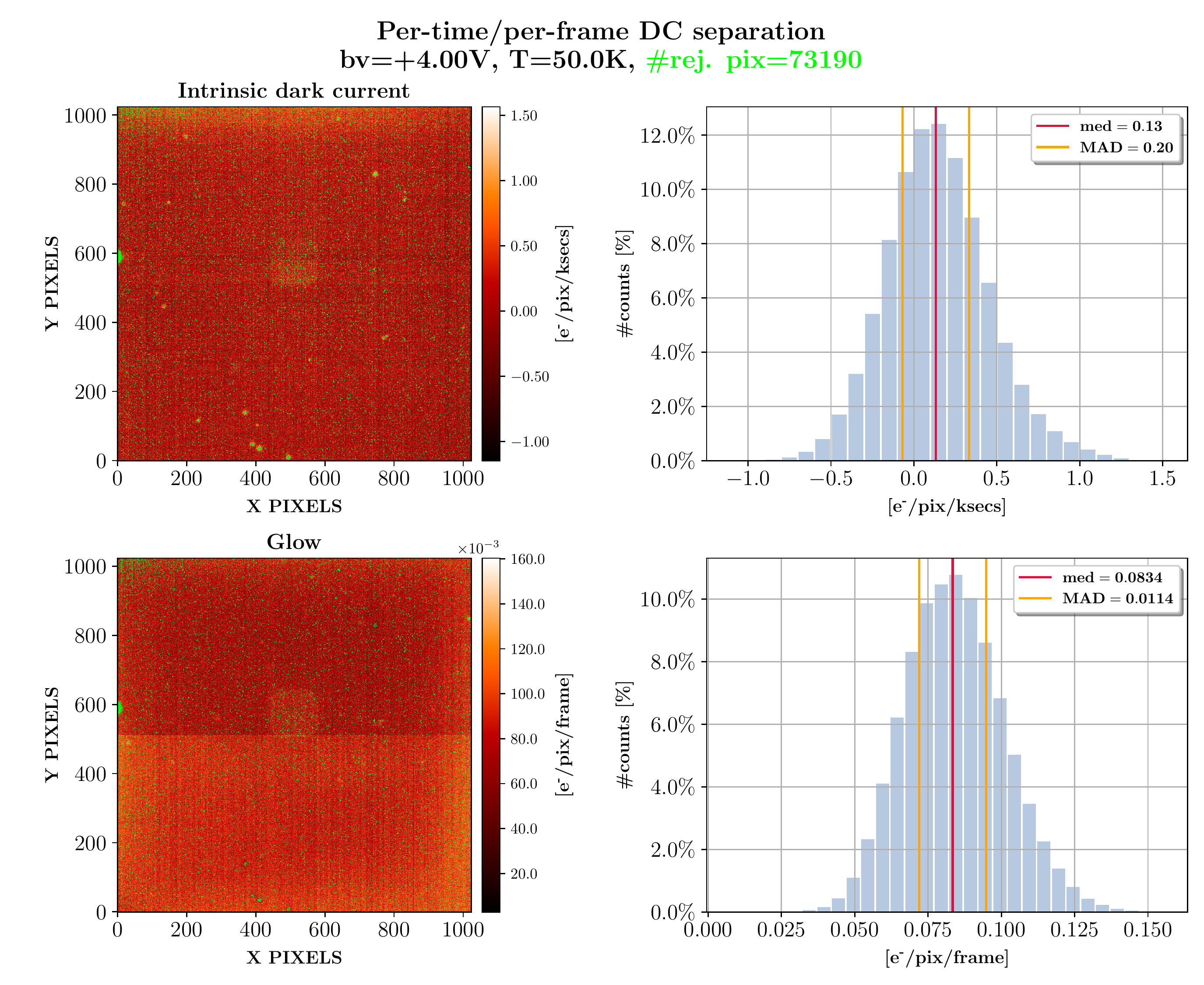}
   \end{tabular}
   \end{center}
   \caption[example] 
   { \label{fig:ptm_pfrm_dark current_sep} 
Heat maps and histograms of the “per-time” or intrinsic dark current (upper row), and “per-frame” glow (lower row). The light green pixels are the pixels rejected as outliers, their total number is given in the plot title. The intrinsic dark current is consistent with zero, with a best estimate of $\sim$0.1 e-/pix/ksec \mbox{(e.g., $\sim$10 e-/pix/day)}, while the glow is consistent with 0.08 e-/pix/frame. The bright “square” in the center is at the location of the hole in our blocking mask, which sees background illumination from the dewar of a few photons per hour. The difference in the top and bottom half of the glow indicates that the glow is higher in the bottom half of the detector, on average 0.015 e-/pix/frame higher. This is thought to be due to the fact that the direct current feeding the electronic components at the origin of this glow is slightly offset in the lower part of the detector.}
   \end{figure}

The results are shown in  Fig.~\ref{fig:ptm_pfrm_dark current_sep}, with the intrinsic dark current of the device is consistent with zero \mbox{(\ttilde0.1 e-/pix/ksec)}, while the glow is \ttilde 0.08 e-/pix/frame. These results are from \ttilde 93\% of active pixels, removing cosmic ray events, hot pixels, and known “problem” areas of the detector (e.g., artifact on the left border). We also repeated this test at 60K for bias voltage=4V, and at 50K for bias voltages=5-8 V, and got consistent results.

The source of the glow photons appears to be within the pixel nodes from the source follower MOSFETs. Previous SAPHIRA designs had the MOSFETs with a different glow shield than the ME1070 ROIC, and reported glow levels two to three orders of magnitude lower\cite{Gert2022}. As such, a future design path is to redo the wafers with the SAPHIRA-style shielding added back in.

There is one important caveat to the dark current results. We identified an increasing number of pixels with anomalously high dark current as we increased our detector bias voltage. The anomalous pixels are primarily comprised of “hot” individual pixels that show dark current several times higher than the background level and are due to the same mechanism as the base tunneling current. These pixel defects were expected to be a problem in these engineering-grade arrays due to a process control issue identified by Leonardo during fabrication, specifically, contamination by silver. As such, the engineering-grade arrays are unsuitable as scientific detectors, but still let us measure the potential of a properly fabricated science-grade array.

\clearpage

\subsection{Overall performance as a function of bias voltage}
\label{sec:perf:results:perfs}

\begin{figure}[ht]
   \begin{center}
   \begin{tabular}{c}
   \includegraphics[scale=0.65]{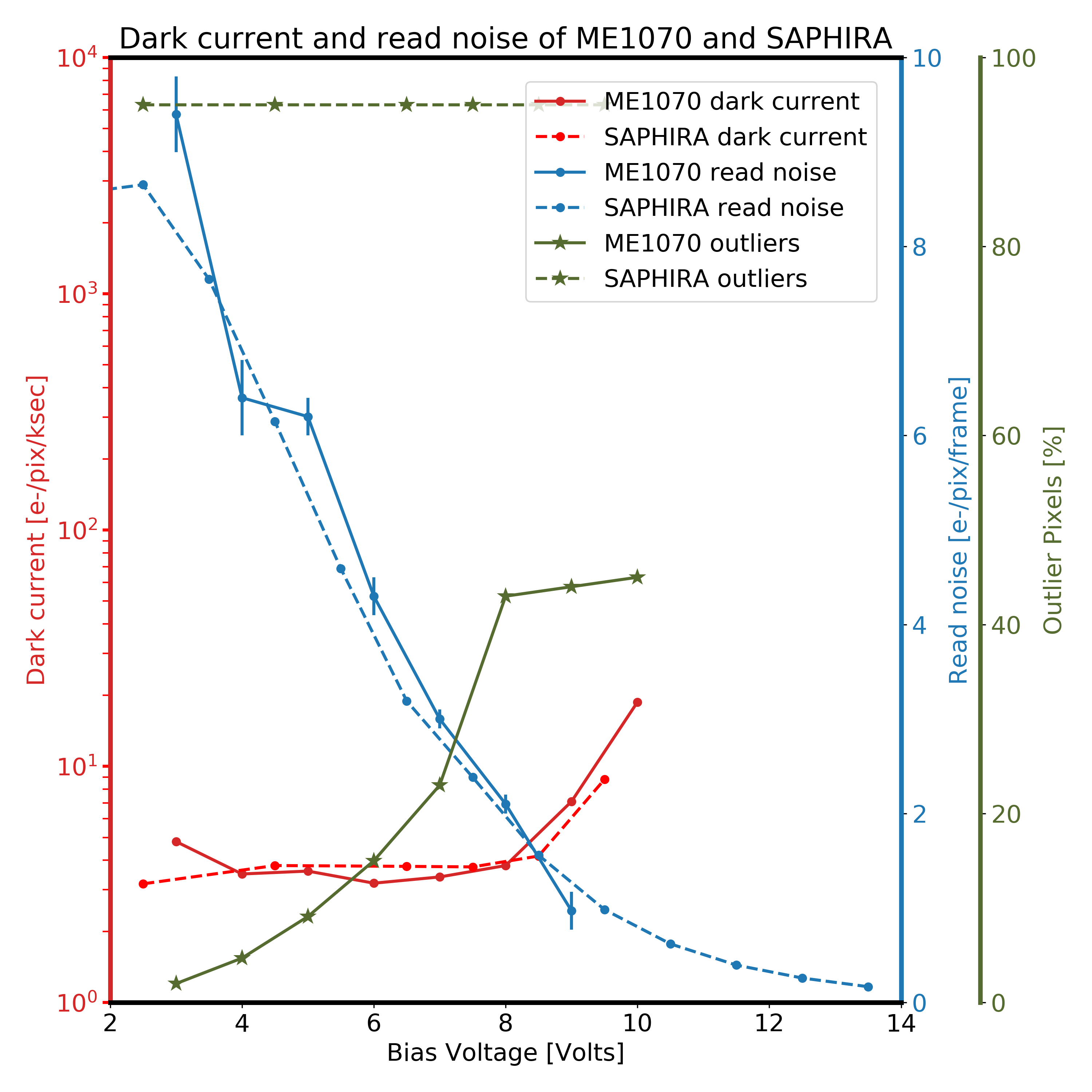}
   \end{tabular}
   \end{center}
   \caption[example] 
   { \label{fig:glob_perfs} 
Measured normalized dark current and effective read noise as a function of bias voltage for the 1kx1k detector and compared to best results obtained with SAPHIRA (data collected from Ref.~\citenum{Atkinson2017,Pastrana2018}). The proportion of pixels rejected as outliers in the dark current tests is also reported. The detector is operated at a temperature of 50K, and set up in source follower mode and with its minimum operating voltages. A steep increase of the dark current starts from bias voltage \textgreater8.0V due to the onset of the tunneling current.}
   \end{figure}

Tab.~\ref{table:glob_perfs} includes all our measurements of conversion gain, read noise and dark current obtained at different bias voltages. Fig.~\ref{fig:glob_perfs} shows the results of our joint analysis of dark current and read noise as a function of bias voltage. They are plotted against the previous best results from the SAPHIRA arrays \cite{Atkinson2017,Pastrana2018}, which have less than 10\% the pixels of the 1kx1k detector. The left axis shows the \textbf{normalized dark current} on a logarithmic scale, and the right axis shows the \textbf{effective read noise} in [e-] on a linear scale. An additional right axis indicates the number of pixels showing anomalous behavior, due to the defects in manufacturing mentioned earlier.

Fig.~\ref{fig:glob_perfs} shows several important effects. First, the effective read noise behaves exactly as expected, reducing by 1.3x per additional volt of bias, consistent with SAPHIRA. The normalized dark current results are also consistent with SAPHIRA and is composed of the glow-limited of \ttilde3 e-/pixel/ksec. This value can be recovered exactly knowing the frame time and per-frame glow of 0.08 e-/pix/frame (0.08e-/pix/frame / (20s/frame) x (1000s/ksec)). As mentioned in the previous section, the intrinsic dark current is consistent with zero. At approximately bias voltage=8V, the tunneling current begins to dominate the dark current budget, and exponentially increases with bias voltage. Finally, we also plot the number of anomalous pixels as a percent of total pixels. In the SAPHIRA devices, the measurements exclude 95\% percent of pixels, as the dark current was measured in a single, isolated region of the device. In the 1kx1k detector, the number of defective pixels is low at bias voltage=4V, but increases approximately quadratically up to $\sim$40\%.

Our engineering-grade devices were first fabricated without optimizing for the onset of tunneling current, so were expected to perform similarly to SAPHIRA with respect to tunneling current offset, at about bias voltage=8V. Theoretical modeling of the device predicted a decrease of about 30\% in read noise per volt. At a starting read noise of $\sim$9 e-/pix/frame, at bias voltage $\sim$3V, we achieve a read noise of $\sim$2 e-/pix/frame at bias voltage $\sim$8V, consistent with theory. The read noise is  reported without the use of any special reduction techniques, e.g., Fowler or MULTIACCUM sampling.\cite{rauscher2007detectors}

% Please add the following required packages to your document preamble:
% \usepackage[table,xcdraw]{xcolor}
% If you use beamer only pass "xcolor=table" option, i.e. \documentclass[xcolor=table]{beamer}
\begin{table}[ht]
\caption{Summary table for different bias voltages of the conversion gain, raw read noise, and effective read noise determined by PTC method, as well as raw dark current and normalized dark current measured using the 2-hour baseline method. The column “Reduction factor" gives the ratio between the conversion gain of the previous bias voltage and the current one. The number of pixels rejected as outliers in the dark current tests is also reported, as well as the effective full well capacity (Eff. FWC). The Eff. FWC is equal to the theoretical maximum dynamic range per pixel of 16 bits, i.e. $2^{16}$ levels in [ADU], multiplied by the conversion gain. The read noise cells for bias voltages=\{8,9,10\}V are shaded in gray to highlight that these PTC measurements of the read noise are less reliable due to noisy data.\newline}
\label{table:glob_perfs}
\resizebox{\textwidth}{!}{%
\begin{tabular}{ccccccccc}
\hline
\textbf{\begin{tabular}[c]{@{}c@{}}Bias\\ voltages {[}V{]}\end{tabular}} & \textbf{\begin{tabular}[c]{@{}c@{}}$\mathbf{K_{gain}}$\\ {[}$\mathbf{\textbf{e}^{-}/\textbf{ADU}}${]}\end{tabular}} & \textbf{\begin{tabular}[c]{@{}c@{}}Reduction\\ factor\end{tabular}} & \textbf{\begin{tabular}[c]{@{}c@{}}Raw RN\\ {[}ADU{]}\end{tabular}} & \textbf{\begin{tabular}[c]{@{}c@{}}Eff. RN\\ {[}$\mathbf{\textbf{e}^{-}}${]}\end{tabular}} & \textbf{\begin{tabular}[c]{@{}c@{}}Raw DC\\ {[}ADU/pix/ks{]}\end{tabular}} & \textbf{\begin{tabular}[c]{@{}c@{}}DC\\ {[}$\mathbf{\textbf{e}^{-}/\textbf{pix}/ks}${]}\end{tabular}} & \textbf{\begin{tabular}[c]{@{}c@{}}\#Rejected\\ pixels {[}\%{]}\end{tabular}} & \textbf{\begin{tabular}[c]{@{}c@{}}Eff. FWC\\ {[}$\mathbf{\textbf{ke}^{-}}${]}\end{tabular}} \\ \hline
\textbf{3.0}                                                             & $1.37 \pm 0.02$                                                                                                     & N.A.                                                                & $6.84 \pm 0.25$                                                     & $9.35 \pm 0.37$                                                                            & $3.5 \pm 0.2$                                                              & $4.8 \pm 0.3$                                                                                         & 2.0                                                                           & 89.6                                                                                         \\
\textbf{4.0}                                                             & $1.02 \pm 0.01$                                                                                                     & 1.34                                                                & $6.24 \pm 0.34$                                                     & $6.37 \pm 0.35$                                                                            & $3.4 \pm 0.9$                                                              & $3.5 \pm 0.9$                                                                                         & 4.7                                                                           & 67.0                                                                                         \\
\textbf{5.0}                                                             & $0.86 \pm 0.005$                                                                                                    & 1.19                                                                & $7.20 \pm 0.21$                                                     & $6.16 \pm 0.18$                                                                            & $4.2 \pm 1.2$                                                              & $3.6 \pm 1.0$                                                                                         & 9.1                                                                           & 56.0                                                                                         \\
\textbf{6.0}                                                             & $0.64 \pm 0.004$                                                                                                    & 1.34                                                                & $6.73 \pm 0.34$                                                     & $4.31 \pm 0.22$                                                                            & $5.0 \pm 1.5$                                                              & $3.2 \pm 1.0$                                                                                         & 15.0                                                                          & 41.9                                                                                         \\
\textbf{7.0}                                                             & $0.49 \pm 0.004$                                                                                                    & 1.31                                                                & $6.23 \pm 0.16$                                                     & $3.03 \pm 0.08$                                                                            & $7.0 \pm 3.0$                                                              & $3.4 \pm 1.5$                                                                                         & 23.0                                                                          & 31.9                                                                                         \\
\textbf{8.0}                                                             & $0.38 \pm 0.003$                                                                                                    & 1.29                                                                & \cellcolor[HTML]{C0C0C0}$5.58 \pm 0.20$                             & \cellcolor[HTML]{C0C0C0}$2.13 \pm 0.08$                                                    & $10.0 \pm 7.0$                                                             & $3.8 \pm 2.7$                                                                                         & 43.0                                                                          & 25.0                                                                                         \\
\textbf{9.0}                                                             & $0.31 \pm 0.004$                                                                                                    & 1.23                                                                & \cellcolor[HTML]{C0C0C0}$3.12 \pm 0.63$                             & \cellcolor[HTML]{C0C0C0}$0.97 \pm 0.20$                                                    & $23.0 \pm 30.0$                                                            & $7.1 \pm 9.3$                                                                                         & 44.0                                                                          & 20.4                                                                                         \\
\cellcolor[HTML]{FFFFFF}\textbf{10.0}                                    & \cellcolor[HTML]{FFFFFF}$0.21 \pm 0.004$                                                                            & \cellcolor[HTML]{FFFFFF}1.48                                        & \cellcolor[HTML]{C0C0C0}$18.4 \pm 0.60$                             & \cellcolor[HTML]{C0C0C0}$3.83 \pm 0.14$                                                    & \cellcolor[HTML]{FFFFFF}$90.0 \pm 200.0$                                   & \cellcolor[HTML]{FFFFFF}$18.7 \pm 41.6$                                                               & 42.0                                                                          & 13.6                                                                                         \\ \hline
\end{tabular}}
\end{table}

\clearpage

\section{OTHER EFFECTS}
\label{sec:other_effects}

While dark current and read noise are the primary objects of this investigation, we have observed several other properties of the detector. During repeated readings under constant illumination, a subset of pixels exhibited what appears to be random telegraph signal (RTS), a spontaneous change of the pixel value between two discrete levels, believed to be caused by random trapping and release of electrons in the readout circuitry of the pixel. In addition, we have evidence of persistence in the detector after bright illumination. We will work to examine RTS and persistence in the coming year. Below, we look at other effects also expected from the engineering-grade devices.

\subsection{Quantum efficiency drop at low bias voltages}
\label{sec:other_effects:QE}

\begin{figure}[ht]
   \begin{center}
   \begin{tabular}{c}
   \includegraphics[scale=0.75]{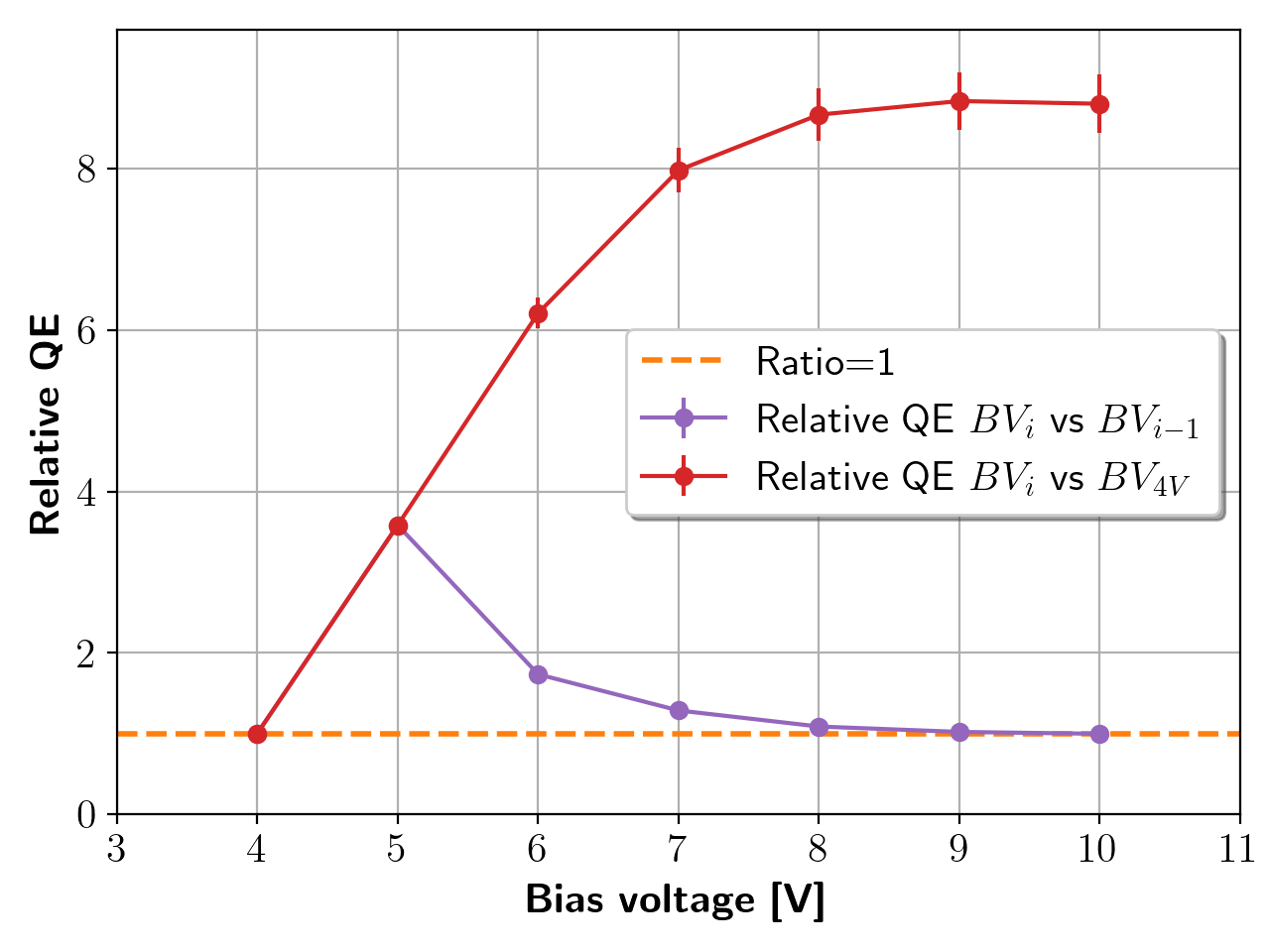}
   \end{tabular}
   \end{center}
   \caption[example] 
   { \label{fig:qe_drop} 
The drop of the relative QE at low bias voltages (BVs) due to contamination in the engineering-grade devices. The purple curve shows the ratio of the QE between two consecutive bias voltages noted $BV_{i}$ and $BV_{i-1}$. The red curve shows the increase of the ratio taking bias voltage=4V as reference. The QE drops by nearly a factor 9 and quasi-linearly from bias voltage=8 to 4V. Beyond bias voltage=8V, the QE reaches a plateau.}
   \end{figure}

The silver contamination issue in the engineering-grade devices impaired the original doping and caused the multiplication region in the detectors to be slightly p-type rather than n-type, resulting in a widening of the gap at the p-n junction. The device needs to be biased sufficiently to overcome this barrier, which effectively causes a steep drop in relative QE at low bias voltages, shown in Fig.~\ref{fig:qe_drop}, with QE dropping by \textgreater80\% from 8 to 4 volts bias. This result was obtained by computing the ratio of the mean signal measured at constant LED flux between  different bias voltages, corrected for the theoretical increase in APD gain, also verified experimentally, as well as for any differences in exposure time.

\subsection{Cosmic rays}
\label{sec:other_effects:CR}

\begin{figure}[ht]
   \begin{center}
   \begin{tabular}{c}
   \includegraphics[scale=0.27]{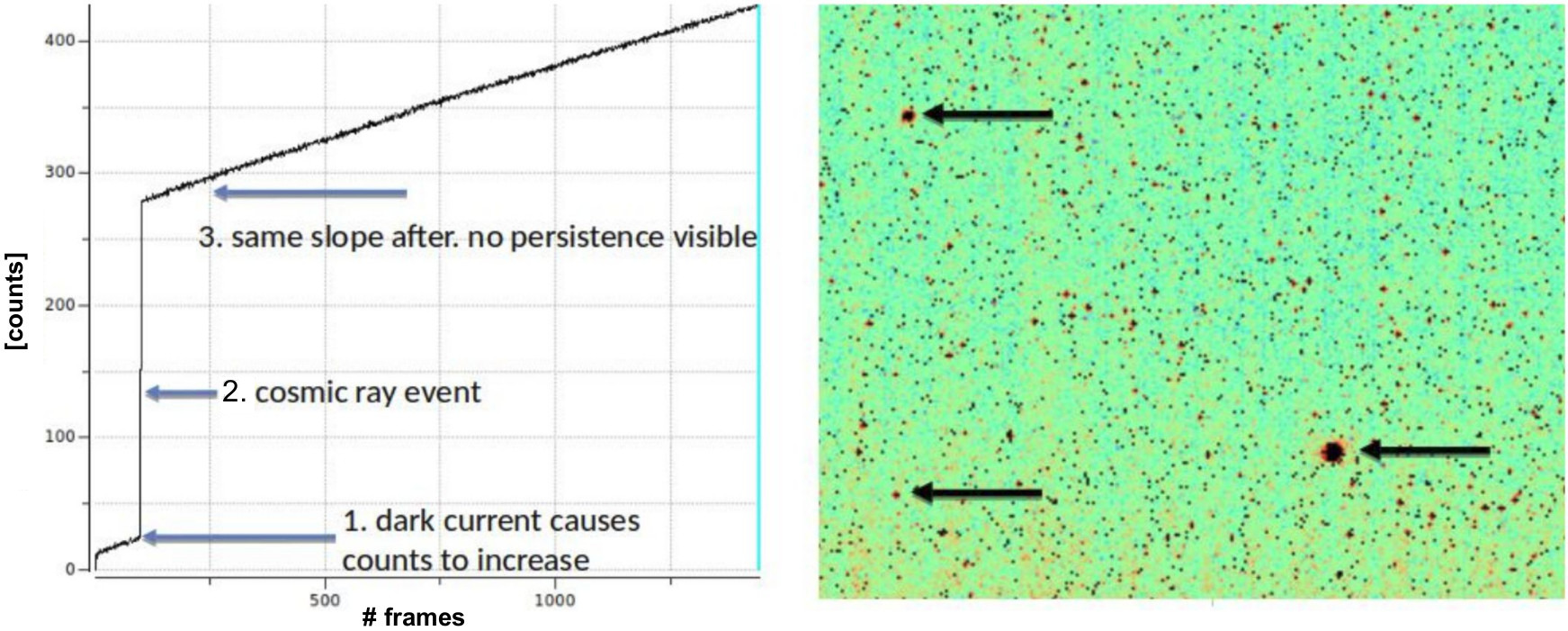}
   \end{tabular}
   \end{center}
   \caption[example] 
   { \label{fig:CR} 
On the left, time series of an individual pixel UTR-sampled at a regular time interval, every 20 seconds, over a total integration time of 8h. The detector is operated in dark conditions, at bias voltage=4.0V, and at a temperature of 50K. At the beginning of the integration, the time series increases with a constant slope of $\sim$0.005 ADU/pix/s, determined by the dark current of this pixel. It exhibits around frame \#100 a large jump in amplitude due to the deposit of charges caused by the impact of a cosmic ray occurring between two consecutive read frames. There is no visible persistence effect on the signal measured afterwards, the slope of the signal returns to its initial evolution prior to the CRE. On the right, a sub-region of the dark current heat map includes three CREs indicated by arrows. A CRE usually affects a cluster of pixels and forms a “hot spot".}
   \end{figure}

Cosmic ray events (CREs) have been detected in our data, especially during long integrations such as those used to measure the dark current. As described in Fig.~\ref{fig:CR}, except for the transient jump visible in the time series of the impacted pixels, no persistence or settling has been observed. The affected pixels are filtered out and rejected as outliers in our tests. We plan to correct for them in future analysis.

\section{CONCLUSION AND PATH FORWARD}
\label{sec:concl}

\subsection{Summary of the results}
\label{sec:concl:sum}

We summarize the above results as follows:
\begin{itemize}
  \item The engineering-grade sensors are providing a large amount of useful data. However, an identified and now solved fabrication issue caused them to have a large amount of defective pixels, and lose QE at low bias voltage. We can filter out these pixels to provide a baseline for the expected performance of the science-grade devices, due to arrive in September.
  \item The intrinsic dark current of these devices is consistent with zero, with a best estimate of $\sim$0.1 e-/pix/ksec. The “dark current” we measure at low bias voltages, of \ttilde 3 e-/pix/ksec, is glow. The glow is measured to be $\sim$0.08 e-/pix/frame, or 1 e-/pix every $\sim$12 frames. The excess glow comes from the source follower MOSFET within the pixel nodes, which has a different metallization shield than the previous generation SAPHIRA devices, which have glow two to three orders of magnitude lower. As such, reverting the metallization to the previous design is a priority for the future.
  \item The read noise of these devices starts around $\sim$10 e-/pix/frame at 3V bias, and reduces by a factor of 1.3 with each additional volt, in agreement with theory.
\end{itemize}

\subsection{Next steps}
\label{sec:concl:next_steps}

The next critical step will be evaluating our science-grade arrays which we expect to begin receiving in September. Initial testing at Leonardo indicates they have several orders of magnitude less defective pixels now that the contamination problem has been addressed. The design of the science grade arrays is also such that the onset of tunneling current should be out at bias voltage \textgreater12 V, rather than 8V, so the dark current will remain low while effective read noise is reduced below 1 e-/pix/frame.

We also plan to examine some effects we observed in our data with the engineering-grade devices, namely random telegraph signal and persistence. Our algorithm for detecting and correcting cosmic ray events will be finalized in order to remove them from our data acquired by UTR sampling during long integration times. Our preliminary estimate of the level of interpixel capacitance will be confirmed with the science-grade sensors using more data to increase the accuracy of our measurements. It would also be valuable to get an independent determination of the charge gain of the detectors by replicating the method by capacitance comparison proposed in Ref.~\citenum{Finger2005}. The unique expected noise performance of these devices holds promise of making them able of photon number resolving in low flux conditions. We plan to conduct experiments to demonstrate this capability.

Following in-lab characterization of the science grade arrays, our next step will be to take some on-sky data. We will be able to use our dewar as an astronomical camera, as it was originally designed. Radiation testing is the final goal. A complete characterization of the detectors will be performed before and after irradiation in order to investigate what effect the radiation has on different device properties and to assess any degradation in performance. This will be a crucial milestone to further advance their qualification for space applications.

\acknowledgments % equivalent to \section*{ACKNOWLEDGMENTS}       
 
The authors would like to thank Dr. Gert Finger of ESO for his helpful comments on our efforts. Development of the detectors and their characterization at UH is sponsored by NASA SAT award \#18-SAT18-0028.

% References
\bibliography{ref} % bibliography data in report.bib
\bibliographystyle{spiebib} % makes bibtex use spiebib.bst

\end{document}